%% file: Template.tex
\documentclass{article}
\usepackage{spconf,amsmath,graphicx,hyperref,amssymb}
\usepackage{tikz}
\usetikzlibrary{shapes.geometric}
\usetikzlibrary{shapes}
\usetikzlibrary{tikzmark, calc,positioning,arrows.meta,quotes}
\usetikzlibrary{calc}
\usetikzlibrary{intersections} 
\usetikzlibrary{backgrounds}
\usetikzlibrary{arrows.meta}
\usepackage{pgfplots}
\usepackage{makecell}
 \usepackage{multirow}
\usepgfplotslibrary{fillbetween}
\usepgfplotslibrary{groupplots}
\pgfplotsset{compat=1.18}
\usepackage{lmodern}
\usepackage{float}
\usepackage{verbatim}
\usepackage{siunitx}
\usepackage{bm}
\usepackage{pgfplots}
\usepackage{silence}
\usepackage{caption}
\usepackage{xcolor}
\usepackage[normalem]{ulem}
\usepackage{booktabs}
\pgfplotsset{compat=1.18} % oder deine installierte Version
\InputIfFileExists{mathmacros}{}{}
\InputIfFileExists{tu_bs_colors}{}{}

\usepackage{orcidlink}

\makeatletter
\ninept
\makeatother

\AtBeginDocument{\normalsize}

\title{Confidence-Aware Speed-over-Ground Estimation \\ from Doppler Spectra\\via Reconstruction-Based Representation Learning}

\name{
\shortstack{
Malte Stelzer$^{\star}$\orcidlink{0009-0003-8875-7664} \quad Marvin Sach$^{\star}$\orcidlink{0000-0003-2909-6539} \quad Timo Bartels$^{\star}$\orcidlink{0009-0002-4917-8849} \quad Dinu Mihailescu-Stoica$^{\dagger}$\orcidlink{0009-0009-5698-7303} \\ 
Sören Kohnert$^{\dagger}$\orcidlink{0009-0005-0625-1198} \quad
Florian Poprawa$^{\dagger}$\orcidlink{0009-0002-2920-4640} \quad Marc Zeller$^{\dagger}$\orcidlink{0000-0002-6738-7903} \quad Tim Fingscheidt$^{\star}$\orcidlink{0000-0002-8895-5041}
}
}

\address{%
\hspace*{-4mm}%
\begin{tabular}{cc}
$^{\star}$Technische Universität Braunschweig, Germany &
$^{\dagger}$Siemens Mobility GmbH, Berlin, Germany \\
\{firstname.lastname, t.fingscheidt\}@tu-bs.de &
\{firstname.lastname\}@siemens.com
\end{tabular}%
}

\begin{document}

\maketitle
\vspace{-2mm}
{\centering\scriptsize
This work has been submitted to the IEEE for possible
publication. Copyright may be transferred without notice,
after which this version may no longer be accessible.\par}
\vspace{-2mm}
\input{sections/0_abstract}

\input{sections/1_Introduction}

\input{sections/3_Method}

\input{sections/4_Experimental_Setup}

\input{sections/5_Experiments}

\input{sections/6_Conclucions}

\bibliographystyle{IEEEbib}
\bibliography{strings,refs}

\end{document}

%% file: mathmacros.tex
%%%%%%%%%%%%%%%%%%%%%%%%%%%%%%%%%%%%%%%%%%%%%%%%%%%%%%%%%%%%%%%%%%%%%%%%%%%%%%%%
% Mathematische Macros
%
% Mathematical macros for lectures and publications
%
% (c) Prof. Dr.-Ing. Tim Fingscheidt
%     IfN / TU Braunschweig, 2006-04-30
%

% Mathematische Zeichen: -------------------------------------------------------

%\newcommand{\conv}{\ast}

   % Residuum
  % Expectation
  % Expectation
%\newcommand{\unit}[1]{{\sf #1}}

         % Quantisierungsfunktion

% SPECOM lecture:
         % Autocorrelation function (ACF)
     % ACF^2
 % ACF vector
    % ACF matrix

   % Crosscorrelation function (CCF)
      % Variance
  % geschaetzte Varianz
     % Standardabweichung
         % pdf

      % Power spectral density (PSD)

          % Vector
     % Vector with Greek letter
          % Matrix
     % Matrix with Greek letter

        % LTP variable (as opposed to LP variable)

%%%%Integrale: El Abed-----------------------------------------------------------

% Anordnen von Indizes usw. unter einer Funktion: ------------------------------
\newcommand{\putindex}[3]{\vtop{\hbox{\hspace{#3} $#1$}
            \hbox{\raise 6mm \hbox{$\scriptscriptstyle #2$}}}}

% Gradient von x, y, allgemein: ------------------------------------------------
\newcommand{\gradx}[0]{\vtop{\hbox{\rm grad}
            \hbox{\raise 2.5mm \hbox{\rm \hspace{2mm} \footnotesize x}}}}

\newcommand{\grady}[0]{\vtop{\hbox{\rm grad}
            \hbox{\raise 2.5mm \hbox{\rm \hspace{2mm} \footnotesize y}}}}

\newcommand{\grad}[1]{\vtop{\hbox{\rm grad}
            \hbox{\raise 2.5mm \hbox{#1}}}}

\newcommand{\btb}{     \begin{tabbing}             }
\newcommand{\bte}{     \end{tabbing}               }

%% file: tu_bs_colors.tex
\definecolor{tu0}{rgb}{0.7451, 0.1176, 0.2353}

\definecolor{tu1}{rgb}{1.0000, 0.8039, 0.0000}
\definecolor{tu11}{rgb}{1.0000, 0.8627, 0.3020}
\definecolor{tu12}{rgb}{1.0000, 0.9020, 0.4980}
\definecolor{tu13}{rgb}{1.0000, 0.9412, 0.6980}
\definecolor{tu14}{rgb}{1.0000, 0.9608, 0.8000}

\definecolor{tu2}{rgb}{0.9804, 0.4314, 0.0000}
\definecolor{tu21}{rgb}{0.9882, 0.6039, 0.3020}
\definecolor{tu22}{rgb}{0.9882, 0.7137, 0.4980}
\definecolor{tu23}{rgb}{0.9922, 0.8275, 0.6980}
\definecolor{tu24}{rgb}{0.9961, 0.8863, 0.8000}

\definecolor{tu3}{rgb}{0.6902, 0.0000, 0.2745}
\definecolor{tu31}{rgb}{0.7529, 0.2000, 0.4196}
\definecolor{tu32}{rgb}{0.8431, 0.4980, 0.6353}
\definecolor{tu33}{rgb}{0.9216, 0.7490, 0.8196}
\definecolor{tu34}{rgb}{0.9529, 0.8510, 0.8902}

\definecolor{tu4}{rgb}{0.4863, 0.8039, 0.9020}
\definecolor{tu41}{rgb}{0.6431, 0.8627, 0.9333}
\definecolor{tu42}{rgb}{0.7412, 0.9020, 0.9490}
\definecolor{tu43}{rgb}{0.8431, 0.9412, 0.9686}
\definecolor{tu44}{rgb}{0.8980, 0.9608, 0.9804}

\definecolor{tu5}{rgb}{0.0000, 0.5020, 0.7059}
\definecolor{tu51}{rgb}{0.3020, 0.6510, 0.7961}
\definecolor{tu52}{rgb}{0.5490, 0.7765, 0.8667}
\definecolor{tu53}{rgb}{0.7490, 0.8745, 0.9255}
\definecolor{tu54}{rgb}{0.8510, 0.9255, 0.9569}

\definecolor{tu6}{rgb}{0.0000, 0.3255, 0.4549}
\definecolor{tu61}{rgb}{0.2510, 0.4941, 0.5922}
\definecolor{tu62}{rgb}{0.5490, 0.6941, 0.7529}
\definecolor{tu63}{rgb}{0.7490, 0.8314, 0.8627}
\definecolor{tu64}{rgb}{0.8510, 0.8980, 0.9176}

\definecolor{tu7}{rgb}{0.7765, 0.9333, 0.0000}
\definecolor{tu71}{rgb}{0.8431, 0.9529, 0.3020}
\definecolor{tu72}{rgb}{0.8863, 0.9647, 0.4980}
\definecolor{tu73}{rgb}{0.9333, 0.9804, 0.6980}
\definecolor{tu74}{rgb}{0.9569, 0.9882, 0.8000}

\definecolor{tu8}{rgb}{0.5373, 0.6431, 0.0000}
\definecolor{tu81}{rgb}{0.6784, 0.7490, 0.3020}
\definecolor{tu82}{rgb}{0.7686, 0.8196, 0.4980}
\definecolor{tu83}{rgb}{0.8588, 0.8941, 0.6980}
\definecolor{tu84}{rgb}{0.9059, 0.9294, 0.8000}

\definecolor{tu9}{rgb}{0.0000, 0.4431, 0.3373}
\definecolor{tu91}{rgb}{0.3020, 0.6118, 0.5373}
\definecolor{tu92}{rgb}{0.5490, 0.7490, 0.7020}
\definecolor{tu93}{rgb}{0.7490, 0.8588, 0.8353}
\definecolor{tu94}{rgb}{0.8549, 0.9176, 0.9059}

\definecolor{tu10}{rgb}{0.8000, 0.0000, 0.6000}
\definecolor{tu101}{rgb}{0.8706, 0.3490, 0.7412}
\definecolor{tu102}{rgb}{0.9216, 0.6000, 0.8392}
\definecolor{tu103}{rgb}{0.9608, 0.8000, 0.9216}
\definecolor{tu104}{rgb}{0.9804, 0.8980, 0.9608}

\definecolor{tu110}{rgb}{0.4627, 0.0000, 0.4627}
\definecolor{tu111}{rgb}{0.5961, 0.2510, 0.5961}
\definecolor{tu112}{rgb}{0.7294, 0.4980, 0.7294}
\definecolor{tu113}{rgb}{0.8392, 0.6980, 0.8392}
\definecolor{tu114}{rgb}{0.9216, 0.8510, 0.9216}

\definecolor{tu120}{rgb}{0.4627, 0.0000, 0.3294}
\definecolor{tu121}{rgb}{0.6118, 0.3020, 0.5333}
\definecolor{tu122}{rgb}{0.7569, 0.5490, 0.6980}
\definecolor{tu123}{rgb}{0.8667, 0.7490, 0.8314}
\definecolor{tu124}{rgb}{0.9216, 0.8510, 0.9020}

\definecolor{tu130}{rgb}{0.0314, 0.0314, 0.0314}
\definecolor{tu131}{rgb}{0.3725, 0.3725, 0.3725}
\definecolor{tu132}{rgb}{0.5882, 0.5882, 0.5882}
\definecolor{tu133}{rgb}{0.7529, 0.7529, 0.7529}
\definecolor{tu134}{rgb}{0.8667, 0.8667, 0.8667}

\definecolor{tu140}{rgb}{0.0000, 0.6875, 0.3125}

%% file: sections/0_abstract.tex
\begin{abstract}
Reliable speed-over-ground estimation from ground-facing Doppler spectra is essential for mobile platforms such as railway systems. Classical signal-processing approaches can provide accurate estimates under favorable conditions, but often degrade in challenging conditions. We propose confidence-aware speed-over-ground estimation trained on rail-domain data that estimates speed over ground and frame-level confidence from Doppler spectra with several novel contributions: First, we transfer the idea of exploiting a reconstruction task by an autoencoder for performance estimation from previous work to Doppler RaDAR. Second, we 
append a speed-over-ground estimation head operating on the autoencoder's latent space features. Third, we
omit the reconstruction decoder during inference but add a low-cost dedicated confidence head for frame-wise error-bound estimation. Fourth, we reveal how noise augmentation training can provide high robustness to the method.  Experiments show that our approach improves speed estimation accuracy over classical baselines while providing informative confidence estimates that indicate the reliability of individual predictions, even under adverse signal conditions.
\end{abstract}

%% file: sections/1_Introduction.tex
\section{Introduction}
\label{sec:Intro}
Ego-velocity estimation is essential in both automotive and railway applications, particularly for automated driving, where highly precise velocity information is required. While automotive systems predominantly rely on inertial measurement units and wheel-speed sensors for ego-velocity estimation \cite{ragab2024automotive}, RaDAR-based approaches have also been investigated in the automotive domain \cite{Shariff2017AdaptiveSpectralEstimation}. In railway applications, sensor configurations include ground-facing Doppler RaDAR, which estimates the train velocity from the Doppler frequency shift in the reflected RaDAR waves from the track bed \cite{reissland2019truespeed}; this task is commonly referred to as speed-over-ground estimation. Reflected RaDAR waves from the train bed can, however,  be affected by surface conditions, weather, contamination, or other environmental influences \cite{reissland2023highaccuracy}.

Although related in terminology, ego-motion and position estimation in robotics constitute a different problem setting and should not be confused with the speed-over-ground estimation considered in this work. Such approaches typically rely on 3D RaDAR data and exploit spatial information such as range, azimuth, and elevation. In contrast, speed-over-ground estimation is based solely on the Doppler spectrum and therefore lacks the explicit 3D spatial information required by these methods \cite{huang2026unrio, zhuang2023iriom, stironja2024rave, kubelka2023scanmatching, abualrub2023radarodometry}, making them not directly applicable to our setting.

Since speed-over-ground estimates can degrade under disturbed measurement conditions, a frame-level confidence measure that reflects the current measurement conditions is needed to identify potentially unreliable predictions, particularly in safety-critical systems \cite{huang2026unrio,sicking2022tailored, geifman2019selectivenet, Fingscheidt2022, 11623919}.

Auxiliary tasks have been shown to improve primary-task performance by encouraging the model to learn more informative shared representations \cite{Chitta2022, Jaeger2023, Stelzer2025}. Reconstruction-based auxiliary tasks are particularly attractive, as the original input provides a directly available reconstruction target and the reconstruction error can serve as an indicator of prediction confidence or uncertainty \cite{baer, Lohdefink_2020_CVPR_Workshops, 9856995}.

In this work, we adapt the idea of \cite{baer} to confidence-aware speed-over-ground estimation from RaDAR Doppler spectra. We transfer the autoencoder-based performance-estimation concept to Doppler RaDAR, but use reconstruction of clean spectra\footnote{ Throughout this work, clean spectra refer to spectra obtained from real RaDAR signals without artificially added noise. Thus, they retain the noise and disturbances inherently present in the measured signals.} \textit{only as an auxiliary training task} and omit the decoder during inference. A \textit{dedicated confidence head} shares encoder features with the new velocity head to provide frame-wise error-bound estimates. We further augment the input spectra with artificial noise to improve robustness under adverse signal conditions. Unlike \cite{baer}, we jointly train the encoder, decoder, and velocity head first, followed by separate optimization of the confidence head.

%% file: sections/3_Method.tex
\section{Method}
\label{sec:method}
\subsection{Architecture}
Our proposed approach for confidence-aware speed-over-ground estimation is illustrated in Figure \ref{fig:diagram}, were we transform an I/Q signal $x(n)$ into a velocity signal $\hat{v}_\ell$ and a corresponding error bound $\hat{\delta}_\ell$ on frame level. We follow the concept introduced in \cite{baer}, but adapt the reconstruction task into an auxiliary training objective and omit it during inference.

\noindent\textbf{I/Q Data and Preprocessing}: Doppler RaDAR signals' I/Q format consists of the in-phase (I) and quadrature (Q) components of the received signal and jointly encode its amplitude and phase information. The complex-valued I/Q signal $x(n) \in \mathbb{C}$, with integer sample index $n \in \mathbb{Z}$, is segmented into frames $\mathbf{x}_\ell \in \mathbb{C}^{N}$, $\ell \in \mathcal{L}=\{1,2,\dots,L\},$ with frame length $N$ using a non-overlapping rectangular window. A discrete Fourier transform (DFT) is then applied to obtain the spectrum $\mathbf{X}_\ell = (X_{\ell,k}) \in \mathbb{C}^{K}$ of frame $\ell$, where $K=2048\geq N$ denotes the number of frequency bins with index $k\in\mathcal{K}=\{0,...,K\!-\!1\}$. If $N<K$, zeros are padded.
\input{figures/architecture_inference}

\noindent\textbf{Spectrum Reconstruction}: The encoder processes the complex DFT spectrum as two real-valued channels, i.e., the real and imaginary parts, with input size $\mathbb{R}^{2 \times K}$. The encoder consists of three 1D convolutional blocks that reduce the spectral dimension from $K=2048$ to $1024$, $512$, and finally $256$, while increasing the channel dimension to $32$, $64$, and $128$. This yields intermediate features $\mathbf{z}'_\ell \in \mathbb{R}^{128 \times 256}$, which retain a position-preserving spectral representation of the input. These features are further compressed by global average pooling and a linear projection into a compact latent bottleneck vector of dimension $\mathbf{z}_\ell \in\mathbb{R}^{128}$. For reconstruction, this bottleneck representation is projected back to a feature map of size $128 \times 256$ and processed by three upsampling convolutional blocks with internal dimensions $128 \times 512$, $64 \times 1024$, and $32 \times 2048$. A final $1 \times 1$ convolution reconstructs the two output channels corresponding to the real and imaginary parts of the reconstructed spectrum $\widehat{\mathbf{X}}_\ell=(\widehat{X}_{\ell,k})\in \mathbb{C}^K$. 

\noindent\textbf{Velocity Estimation}: The velocity head applies an additional 1D convolutional block to $\mathbf{z}'_\ell$, followed by dropout and a $1 \times 1$ convolution that reduces the channel dimension to one. The resulting one-dimensional feature sequence is flattened and mapped by a final linear layer to the sequence of scalar velocity estimates $\hat{v}_\ell$, $\ell\in\mathcal{L}$.  

\noindent\textbf{Confidence Estimation}: The confidence head is architecturally identical to the velocity head, with the only difference being a Softplus activation function \cite{dugas2001incorporating} at the output to ensure non-negative predictions. It estimates an upper error bound $\hat{\delta}_\ell$ corresponding to the 95th percentile of the absolute velocity error. \textit{Thus, for a well-calibrated model, approximately 95\% of the absolute velocity errors are expected to lie below the predicted error bound $\hat{\delta}_\ell$.} Since smaller values of $\hat{\delta}_\ell$ indicate a tighter expected error bound, they correspond to higher confidence in the velocity estimate.

While the overall processing pipeline is inspired by \cite{baer}, we employ a more compact network architecture than image-based approaches, since our inputs are one-dimensional Doppler spectra and our targets comprise a reconstructed spectrum and a scalar velocity, resulting in substantially lower input and output dimensionality. We further incorporate a dedicated module for confidence estimation and get rid of the comutationally demanding compression and reconstruction decoding in inference ($\sim$ 10 times less FLOPs).
\subsection{Training}
\label{ssec:train}
\noindent\textbf{Reference Velocity}: The reference velocity $v_\ell$ is created by fusing odometrical pulse generators (OPGs) mounted on two wheel axels of a wagon together with a tight coupling (global navigation satellite system (GNSS)/inertial measurement unit (IMU)) localization system and an optical speed correlation sensor. The wheel diameters are auto-calibrated via the GNSS velocity. To detect and exclude phases containing slip and slide effects\footnote{\textit{Slip and slide} describe deviations between the wheels' rotation speed and the true vehicle speed caused by limited wheel-rail adhesion during acceleration or braking.} of the OPGs, their velocity estimates are compared to each other and to the velocities provided by GNSS and the optical sensor. The optical sensor improves the fused velocity estimate to reference-level accuracy; however, this processing is not available at runtime.

\noindent\textbf{Losses}: For training our network, we employ a loss function consisting of two components. To learn to reconstruct the spectrum, we use as first loss component \cite{braun}
\begin{align}
J_\ell^\mathrm{spec}
&=
\tfrac{1}{K}
\textstyle\sum\nolimits_{k\in\mathcal{K}}
(1-\beta)\cdot(
|\widehat{X}_{\ell,k}|^{c}
-
|\overline{X}_{\ell,k}|^{c}
)^2
\notag
\\
&\quad
+\beta\cdot ||
|\widehat{X}_{\ell,k}|^{c}
e^{j\hat{\varphi}_{\ell,k}}
-
|\overline{X}_{\ell,k}|^{c}
e^{j\overline{\varphi}_{\ell,k}}
||^2,
\label{eq:spec}
\end{align}
where $\overline{\varphi}_{\ell,k}$ and $\hat{\varphi}_{\ell,k}$ denote the phases of $\overline{X}_{\ell,k}$ and $\widehat{X}_{\ell,k}$, respectively. Note that target spectra hold $\overline{X}_{\ell,k}=X_{\ell,k}$. Furthermore, $\beta \in [0,1]$ is a weighting factor, and $c$ denotes an amplitude compression factor. The second loss component is the mean absolute error (MAE)
\begin{equation}
J_\ell^\mathrm{vel}
=
\left|\hat{v}_\ell-v_\ell\right|
\label{eq:vel_loss}
\end{equation}
between the estimated and reference velocities. Both loss components are combined into the overall loss function
\begin{equation}
J_\ell
=
(1-\alpha)J_\ell^\mathrm{spec}
+
\alpha J_\ell^\mathrm{vel},
\label{eq:loss}
\end{equation}
where $\alpha \in [0,1]$ balances the spectral reconstruction and velocity estimation losses. Unlike \cite{baer}, we do not initialize a two-stage procedure with velocity estimation, as this simpler task could yield a less informative $\mathbf{z}'_\ell$ and limit reconstruction performance. Instead, the encoder, reconstruction decoder, and velocity head are trained jointly first, then frozen while only the confidence head is optimized using the pinball loss \cite{koenker1978regression}
\vspace*{-5mm}
\begin{equation}
J_\ell^{\mathrm{conf}}
=
\begin{cases}
\tau \left(\left|\hat{v}_\ell-v_\ell\right|-\delta_\ell\right),
& \left|\hat{v}_\ell-v_\ell\right| \geq \delta_\ell, \\[1mm]
(1-\tau)\left(\delta_\ell-\left|\hat{v}_\ell-v_\ell\right|\right),
& \left|\hat{v}_\ell-v_\ell\right| < \delta_\ell,
\end{cases}
\label{eq:pinball_loss}
\end{equation}
with the quantile $\tau\in [0,1]$. Learning rate, weight decay, and data are same as in the first stage.

\noindent\textbf{Training Noise}: To make spectrum reconstruction and velocity estimation more robust against degraded RaDAR spectra, we also add artificial noise to the spectra $\mathbf{X}_{\ell}$ during training. The signal power in frame $\ell$ is estimated as
\begin{equation}
P^{\mathrm{(x)}}_\ell
=
\tfrac{1}{K}
\sum_{k\in\mathcal{K}}
\left|X_{\ell,k}\right|^2
\label{eq:snr_}
\end{equation}
For a chosen signal-to-noise
ratio $\mathrm{SNR}$ (in dB), we then obtain the noise power
\begin{equation}
P^{\mathrm{(\epsilon)}}_\ell
=
\tfrac{P^{\mathrm{(x)}}_\ell}
{10^{\mathrm{SNR}/10~\mathrm{dB}}}.
\label{eq:snr}
\end{equation}
The complex-valued noise $\bm\epsilon_\ell=(\epsilon_{\ell,k})$ with $\epsilon_{\ell,k}\sim\mathcal{N}_\mathbb{C}(0,P_\ell^{\mathrm{(\epsilon)}})$ is drawn from a complex normal distribution and then added independently in
each time-frequency bin as
\begin{equation}
\widetilde{X}_{\ell,k}
=
X_{\ell,k}
+
\epsilon_{\ell,k},
\label{eq:noise}
\end{equation}
with $(\tilde{X}_{\ell,k})=\tilde{\mathbf{X}}_\ell$. This frequency-domain noise model is equivalent to adding circularly symmetric complex white Gaussian noise before the DFT, up to a constant factor. The noise is not added to the targets $\overline{\mathbf{X}}_\ell = \mathbf{X}_\ell$ of the reconstruction, only to the input $\mathbf{X}_\ell$ of the encoder. The SNRs and noise schedule used in training are described in Section \ref{sec:experimental}.

%% file: figures/architecture_inference.tex
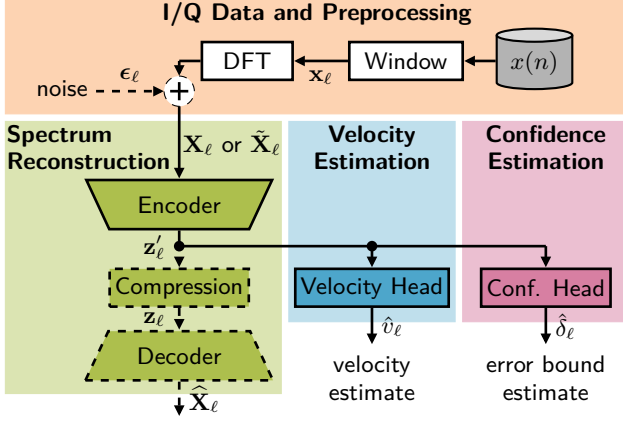
\begin{figure}[t!]
\centering

        \begin{tikzpicture}[node distance=2cm, auto, font=\sffamily]
            \tikzstyle{arrow} = [-{Triangle[length=4.1pt,width=3.7pt]}, line width=1pt]
            \tikzstyle{line} = [ line width=1pt]
            \tikzstyle{arrow_d} = [-{Triangle[length=4.1pt,width=3.7pt]}, line width=1pt, dashed]
            \tikzstyle{line_d} = [ line width=1pt, dashed]

                % Rectangle IQ
                \draw [fill=tu23, tu23] (0,0) rectangle (8.25,-1.5);

                % Rectangle Reconstruction
                \draw [fill=tu83, tu83] (0,-1.6) rectangle (3.65,-5.2);

                % Rectangle Velocity Head
                \draw [fill=tu53, tu53] (3.75,-1.6) rectangle (5.95,-4.25);

                % Rectangle Velocity Head
                \draw [fill=tu33, tu33] (6.05,-1.6) rectangle (8.25,-4.25);
                
                % IQ
                \node[cylinder, draw, thick, aspect = 0.8, minimum width = 1cm, minimum height = 0.8cm, align=center,  rotate=90, cylinder uses custom fill, cylinder body fill=gray!60!white , cylinder end fill=gray!30!white ] (IQ) at (7,-0.85) {};

                \node[anchor=south, align= center] at (7,-1.1) {$x(n)$};
                \node[anchor=south, align= center] at (4.125,-0.45) {\textbf{I/Q Data and Preprocessing}};

                \node[
                          rectangle,
                          draw,
                          line width=1pt,
                          fill=white,
                          minimum height=0.5cm,
                          minimum width=1.5cm,
                          inner xsep=2pt
                        ] (window) at (5.3,-0.8) {Window};

                \node[
                          rectangle,
                          draw,
                          line width=1pt,
                          fill=white,
                          minimum height=0.5cm,
                          minimum width=1.2cm,
                          inner xsep=2pt
                        ] (dft) at (3.2,-0.8) {DFT};

                \draw[arrow] ([yshift=0.5mm]IQ.north) -- (window.east) ;
                \draw[arrow] (window.west) -- (dft.east) node[midway, below] {$\mathbf{x}_\ell$};

\begin{scope}[shift={(2.3,-1.2)}]
    \fill[white] (0,0) circle (2mm);
    \draw[dashed] (0,0) circle (2mm);
    \draw[line width=1pt] (-1.2mm,0) -- (1.2mm,0);
    \draw[line width=1pt] (0,-1.2mm) -- (0,1.2mm);
\end{scope}

                % Reconstruction
                \node[anchor=south, align= left] at (1.10,-2.35) {\textbf{Spectrum}\\\textbf{Reconstruction}};

                \node[
                          trapezium,
                          trapezium angle=65,
                          rotate=180,
                          draw,
                          line width=1pt,
                          fill=tu81,
                          minimum height=0.65cm,
                          minimum width=0.8cm,
                          inner xsep=2pt
                        ] (enc) at (2.3,-2.7) {\rotatebox{180}{Encoder}};

                \draw[line] (dft.west) -- (2.3,-0.8) ;
                \draw[arrow] (2.3,-0.8) -- (2.3,-1.0);
                \draw[arrow] (2.3,-1.4) -- (enc.south) node[midway, yshift=0mm,xshift=-0.8mm] {$\mathbf{X}_\ell$ or $\tilde{\mathbf{X}}_\ell$};
                \draw[arrow, dashed] (1.2,-1.2) -- (2.1,-1.2)node[midway] {$\bm{\epsilon}_\ell$};
                \node[anchor=south, align= center] at (0.75,-1.38) {noise};

                \node[
                          rectangle,
                          draw, dashed,
                          line width=1pt,
                          fill=tu81,
                          minimum height=0.5cm,
                          minimum width=1.5cm,
                          inner xsep=2pt
                        ] (comp) at (2.3,-3.8) {Compression};

                \node[
                          trapezium,
                          trapezium angle=65,
                          draw, dashed,
                          line width=1pt,
                          fill=tu81,
                          minimum height=0.65cm,
                          minimum width=0.8cm,
                          inner xsep=2pt
                        ] (dec) at (2.3,-4.7) {Decoder};

                \draw[line] (enc.north) -- (2.3, -3.2);
                \draw[arrow, dashed] (2.3, -3.2) -- (comp.north) node[midway, yshift=1mm, xshift=-6mm] {$\mathbf{z}'_\ell $};
               % \draw[arrow, dashed] (enc.north) -- (comp.north) node[midway, xshift=-6mm] {$\mathbf{z}'_\ell $};

                \draw[arrow, dashed] (comp.south) -- (dec.north) node[midway, xshift=-6mm] {$\mathbf{z}_\ell $};

                \draw[arrow, dashed] (dec.south) -- node[midway] {$\widehat{\mathbf{X}}_\ell$} (2.3,-5.5);

                \fill (2.3,-3.25) circle (2pt);

                % Velocity Estimation
                \node[anchor=south, align= center] at (4.85,-2.35) {\textbf{Velocity}\\\textbf{Estimation}};

                \node[
                          rectangle,
                          draw,
                          line width=1pt,
                          fill=tu51,
                          minimum height=0.5cm,
                          minimum width=1.5cm,
                          inner xsep=2pt
                        ] (velocity) at (4.85,-3.8) {Velocity Head};

                \draw[line] (2.3,-3.25) -- (4.85,-3.25);
                \draw[arrow] (4.85,-3.25) -- (velocity.north);
                \draw[arrow] (velocity.south) -- node[midway] {$\hat{v}_\ell$} (4.85,-4.55);
                \node[anchor=south, align= center] at (4.85,-5.4) {velocity\\estimate};

                \node[anchor=south, align= center] at (7.15,-2.35) {\textbf{Confidence}\\\textbf{Estimation}};

                \node[
                          rectangle,
                          draw,
                          line width=1pt,
                          fill=tu32,
                          minimum height=0.5cm,
                          minimum width=1.5cm,
                          inner xsep=2pt
                        ] (velocity) at (7.15,-3.8) {Conf. Head};

                \draw[line] (4.85,-3.25) -- (7.15,-3.25);
                \draw[arrow] (7.15,-3.25) -- (velocity.north);
                \draw[arrow] (velocity.south) -- node[midway]{$\hat{\delta}_\ell$} (7.15,-4.55)  ;
                \fill (4.85,-3.25) circle (2pt);

                \node[anchor=south, align= center] at (7.15,-5.4) {error bound\\estimate};

        \end{tikzpicture}
\vspace*{-2mm}
\caption{\textbf{Our proposed speed-over-ground estimation with confidence estimation}. Inference: Solid signal paths and blocks; training: also dashed ones.}
\vspace*{-2mm}
\label{fig:diagram}
\end{figure}

%% file: sections/4_Experimental_Setup.tex
\begin{table*}[ht!] 
\centering 
\caption{\textbf{MAE results on} $\mathcal{D}^\mathrm{val}$ \textbf{and} $\mathcal{D}^\mathrm{test}$: Average velocity MAE in m/s across different noise conditions. For our models, we investigated noise augmentation training and vary $\alpha$ in (\ref{eq:loss}). Best values are marked bold, second-best are underlined. } 

\newcolumntype{E}{%
    r@{\hspace{0.05em}}%
    >{\raisebox{0.7ex}\bgroup\scriptsize}r<{\egroup}%
}
\setlength{\tabcolsep}{2pt}
\label{tab:results} 
\begin{tabular}{
l|c|c|
r r r|
r
r
r
r|
c}
\toprule
Method
& \makecell{Train.\\noise}
& $\alpha$
& \multicolumn{1}{c}{\makecell{$\mathcal{D}^\mathrm{val}$\\clean}}
& \multicolumn{1}{c}{\makecell{$\mathcal{D}^\mathrm{val}$\\+ noise}}
& \multicolumn{1}{c|}{\makecell{$\mathcal{D}^\mathrm{val}$\\+ -10 dB noise}}
& \multicolumn{1}{c}{\makecell{$\mathcal{D}^\mathrm{test}$\\clean}}
& \multicolumn{1}{c}{\makecell{$\mathcal{D}^\mathrm{test}$\\+ noise}}
& \multicolumn{1}{c}{\makecell{$\mathcal{D}^\mathrm{test}$\\+ -5 dB noise}}
& \multicolumn{1}{c}{\makecell{$\mathcal{D}^\mathrm{test}$\\+ -10 dB noise}}
& \multicolumn{1}{|c}{\makecell{$\mathcal{D}^\mathrm{test}$\\avg.\ rank $\downarrow$}}
\\
\midrule
\midrule \texttt{PP} \cite{richards2010principles} 
& - 
& -
& ${0.41}^{\pm0.47}$
& ${0.45}^{\pm0.06\phantom{1}}$
& ${1.02}^{\pm11.24}$
& ${0.42}^{\pm0.43}$
& ${0.45}^{\pm0.07\phantom{1}}$ 
& ${0.60}^{\pm6.34\phantom{1}}$
& ${1.17}^{\pm12.87}$
& 6.00\\

 \texttt{CMA} \cite{Shariff2017AdaptiveSpectralEstimation} 
 & - 
 & - 
 & ${1.04}^{\pm1.73}$
 & ${1.30}^{\pm1.54\phantom{1}}$
 & ${24.68}^{\pm68.23}$
 & ${0.99}^{\pm1.51}$
 & ${1.17}^{\pm1.31\phantom{1}}$
 & ${4.09}^{\pm26.56}$
 & ${25.07}^{\pm69.00}$
 & 7.00\\
 
 \texttt{XCA} \cite{Shariff2017AdaptiveSpectralEstimation} 
 & - 
 & - 
 & ${0.36}^{\pm1.41}$
 & ${0.36}^{\pm0.03\phantom{1}}$
 & ${\underline{0.38}}^{\pm1.52\phantom{1}}$
 & ${0.34}^{\pm1.31}$
 & ${0.34}^{\pm0.00\phantom{1}}$
 & ${0.35}^{\pm1.39\phantom{1}}$
 & ${\underline{0.37}}^{\pm1.59\phantom{1}}$
 & 3.50\\\midrule

 \multirow{10}{*}{Ours}

& no
& 1.0
& ${\textbf{0.14}}^{\pm0.20}$
& ${9.62}^{\pm7.91\phantom{1}}$
& ${33.63}^{\pm18.92}$
& ${\textbf{0.15}}^{\pm0.20}$
& ${9.63}^{\pm7.91\phantom{1}}$
& ${23.68}^{\pm13.87}$
& ${33.93}^{\pm19.28}$
& 6.75\\

& no 
& 0.9
& ${\textbf{0.14}}^{\pm0.20}$
& ${13.02}^{\pm12.42}$
& ${37.57}^{\pm20.08}$
& ${\textbf{0.15}}^{\pm0.19}$ 
& ${13.18}^{\pm12.54}$ 
& ${35.94}^{\pm19.10}$
& ${38.17}^{\pm20.93}$
& 8.25\\

& no 
& 0.8 
& ${\underline{0.15}}^{\pm0.21}$
& ${11.01}^{\pm11.99}$
& ${36.70}^{\pm19.50}$
& ${\textbf{0.15}}^{\pm0.22}$ 
& ${11.16}^{\pm12.14}$ 
& ${34.40}^{\pm18.66}$
& ${37.20}^{\pm20.17}$ 
& 7.50\\

& yes 
& 1.0
& ${\underline{0.15}}^{\pm0.19}$
& ${\textbf{0.18}}^{\pm0.02\phantom{1}}$ 
& ${\mathbf{0.37}}^{\pm1.00\phantom{1}}$
& ${\textbf{0.15}}^{\pm0.20}$ 
& ${\textbf{0.18}}^{\pm0.02\phantom{1}}$ 
& ${\mathbf{0.21}}^{\pm0.44\phantom{1}}$
& ${\textbf{0.36}}^{\pm0.99\phantom{1}}$ 
& \textbf{1.00}\\

& yes 
& 0.9 
& ${\textbf{0.14}}^{\pm0.18}$
& ${\textbf{0.18}}^{\pm0.02\phantom{1}}$ 
& ${0.49}^{\pm1.27\phantom{1}}$
& ${\textbf{0.15}}^{\pm0.21}$ 
& ${\textbf{0.18}}^{\pm0.02\phantom{1}}$ 
& ${\underline{0.22}}^{\pm0.41\phantom{1}}$
& ${0.46}^{\pm1.10\phantom{1}}$
& 2.50\\

& yes 
& 0.8 
& ${\underline{0.15}}^{\pm0.20}$
& ${\textbf{0.18}}^{\pm0.03\phantom{1}}$
& ${0.50}^{\pm1.36\phantom{1}}$
& ${\textbf{0.15}}^{\pm0.21}$ 
& ${\underline{0.19}}^{\pm0.03\phantom{1}}$ 
& ${0.23}^{\pm0.47\phantom{1}}$
& ${0.48}^{\pm1.22\phantom{1}}$
& 3.50\\

& yes
& 0.7
& ${\underline{0.15}}^{\pm0.18}$
& ${\textbf{0.18}}^{\pm0.02\phantom{1}}$
& ${0.44}^{\pm1.04\phantom{1}}$
& ${\textbf{0.15}}^{\pm0.20}$ 
& ${\underline{0.19}}^{\pm0.02\phantom{1}}$ 
& ${0.23}^{\pm0.51\phantom{1}}$
& ${0.43}^{\pm1.09\phantom{1}}$
& 2.75\\

& yes
& 0.6
& ${\underline{0.15}}^{\pm0.18}$
& ${\underline{0.19}}^{\pm0.03\phantom{1}}$
& ${0.41}^{\pm0.98\phantom{1}}$
& ${\underline{0.16}}^{\pm0.26}$ 
& ${\underline{0.19}}^{\pm0.02\phantom{1}}$ 
& ${0.23}^{\pm0.43\phantom{1}}$
& ${0.40}^{\pm0.90\phantom{1}}$
& 2.50\\

& yes
& 0.5
& ${\underline{0.15}}^{\pm0.20}$
& ${\textbf{0.18}}^{\pm0.02\phantom{1}}$
& ${0.43}^{\pm1.27\phantom{1}}$
& ${\textbf{0.15}}^{\pm0.20}$ 
& ${\underline{0.19}}^{\pm0.03\phantom{1}}$ 
& ${\underline{0.22}}^{\pm0.46\phantom{1}}$
& ${0.41}^{\pm1.18\phantom{1}}$
& \underline{2.25}\\

& yes
& 0.4
& ${\underline{0.15}}^{\pm0.20}$
& ${\underline{0.19}}^{\pm0.03\phantom{1}}$
& ${0.50}^{\pm1.20\phantom{1}}$
& ${\underline{0.16}}^{\pm0.20}$ 
& ${0.20}^{\pm0.03\phantom{1}}$ 
& ${0.24}^{\pm0.46\phantom{1}}$
& ${0.47}^{\pm1.12\phantom{1}}$
& 4.00\\

\bottomrule 
\end{tabular} 
\end{table*}

\section{Experimental Setup}
\label{sec:experimental}
\noindent\textbf{Dataset}: The acquisition chain uses a continuous-wave patch antenna at $f_c=24$ GHz. The ground-reflected RF signal is mixed with the transmitted signal and a 90° phase-shifted copy to obtain the in-phase (I) and quadrature (Q) components. The I/Q imbalance is below 2\% and neglected. Both signals are high-pass filtered to remove biases, low-pass filtered for anti-aliasing, and A/D-converted at $f_s=83.682$ kHz. The antenna is tilted by $\vartheta=\ang{37}$ relative to the horizontal and has a half-power beamwidth of 5°.

We split the data into a training set $\mathcal{D}^\mathrm{train}$ with $200\mathrm{k}$ frames, a validation set $\mathcal{D}^\mathrm{val}$ with $20\mathrm{k}$ frames, and a test set $\mathcal{D}^\mathrm{test}$ with $20\mathrm{k}$ frames. Note that each set contains frames from disjoint routes, thereby covering different weather and ground conditions. The training set comprises 20 routes, while the validation and test sets each comprise 10 routes. Frames were assigned such that the velocity distributions are as similar as possible across the sets. 

\noindent\textbf{Validation/Test Noise}: To evaluate the robustness of the velocity estimators under noisy conditions, additive noise is applied also for validation and test purposes to the complex spectra of the data (\ref{eq:noise}). Signal power estimation follows (\ref{eq:snr_}).

\noindent\textbf{Metrics}: As the primary evaluation metric, we use the MAE between $\hat{v}_\ell$ and $v_\ell$, consistent with \eqref{eq:vel_loss}. It is reported on the validation and test sets for clean and noisy data, both per noise level and averaged across all considered noise levels. MAE and standard deviation are given in m$/$s.

\noindent\textbf{Parameterization}:
For our experiments, we use a frame length of $N=1287$, a compression factor of $c=0.3$ following \cite{braun}, and a reconstruction weight of $\beta=0.3$. Noise is introduced via a curriculum schedule with probabilities of 0.3, 0.5, and 0.7 and SNR ranges of $[15, 25]$, $[5, 20]$, and $[-5, 15]$ dB starting from epochs 11, 26, and 41, respectively. The network is trained for 80 epochs with an initial learning rate of $10^{-3}$, halved every 15 epochs, weight decay of $10^{-4}$, velocity-head dropout of 0.1, and quantile $\tau=0.95$.

\noindent\textbf{Reference Methods}:
As classical baselines, we consider three frequency-domain estimators. Peak picking (\texttt{PP}) provides the most direct approach by selecting the frequency bin $k_\ell^\mathrm{max}$ with maximum spectral magnitude \cite[p.~654]{richards2010principles}, while ignoring seven bins around DC to suppress residual low-frequency bias. Then, $v_\ell= \frac{\left(k_\ell^\mathrm{max}-\frac{K}{2}\right)\frac{f_s}{K}c_0}{2f_c\cos(\vartheta)}$, where $c_0$ is the speed of light in vacuum. Center of mass (\texttt{CMA}) \cite{Shariff2017AdaptiveSpectralEstimation} estimates the Doppler frequency from the spectral mass within an adaptively selected signal band, reducing sensitivity to isolated spectral peaks. Cross-correlation (\texttt{XCA}) \cite{Shariff2017AdaptiveSpectralEstimation} smoothes the Doppler spectrum, constructs a Gaussian reference from the coarse peak estimate and antenna beamwidth, and selects the frequency shift that maximizes the correlation with this template.

%% file: sections/5_Experiments.tex
\section{Experimental Validation and Discussion}
\label{sec:experiments}
Table \ref{tab:results} summarizes performance on $\mathcal{D}^\mathrm{val}$ and $\mathcal{D}^\mathrm{test}$ and compares our method with the reference methods. We evaluate different $\alpha$ in (\ref{eq:loss}) and additive noise training (\ref{eq:noise}); $\alpha=1.0$ disables reconstruction training. Mean and standard deviation of the MAE are reported. For ''+ noise'', both are computed across SNRs of $-5$, $0$, $5$, $10$, $15$, and $20$ dB; for clean data and individual SNRs, they are computed across samples. The average rank is also reported on $\mathcal{D}^\mathrm{test}$.

On $\mathcal{D}^\mathrm{val}$, we observe that \texttt{XCA} \cite{Shariff2017AdaptiveSpectralEstimation} is the best among all baseline methods, with MAE only slightly depending on SNR. Our approaches are top in clean conditions, however, with a severe degradation in noisy conditions. Our method \textit{with} noise augmentation training, however, very much improves in both noise conditions. Among our configurations, we choose $\alpha = 1.0$, as it delivers two 1st ranks and one 2nd on $\mathcal{D}^\mathrm{val}$.

On $\mathcal{D}^\mathrm{test}$, the overall ranks are very similar to $\mathcal{D}^\mathrm{val}$: \texttt{XCA} \cite{Shariff2017AdaptiveSpectralEstimation} again is best among baselines and 2nd at very harsh noise ($-10$ dB). Our approaches with noise augmentation training show overall top performance in all four conditions. \textit{Our method with $\alpha=1.0$, i.e., without reconstruction loss (\ref{eq:spec}) in (\ref{eq:loss}), secures the first rank among all from clean data unto $-10$ dB test noise.} The ranking shows another strong result for $\alpha\in \{0.5, 0.6\}$, achieving the second-best rank. So far, we consider SNRs below $-10$ dB as out-of-operation range, errors increase above 0.4 m/s. Checking robustness towards $-15$ dB, however, the noise-augmented model with $\alpha=0.6$ achieves an MAE of 2.23 m/s versus 4.30 m/s for $\alpha=1.0$.

Figure \ref{fig:velocity} shows the velocity MAE of the three reference methods and our selected best-performing configuration across SNRs ranging from $-10$ to $20$ dB, therefore illustrating the robustness and quality of the velocity prediction under different noise conditions. Our method shows almost constant MAE of about $0.15$ m/s (0.54 km/h) over the entire SNR range unto $-5$ dB. At $-10$ dB, it degrades to the level of \texttt{XCA} \cite{Shariff2017AdaptiveSpectralEstimation}(MAE of 0.34 m/s). In contrast, the performance of \texttt{PP} \cite{richards2010principles} deteriorates markedly below $5$ dB, indicating a substantially higher sensitivity to noise. \texttt{CMA} \cite{Shariff2017AdaptiveSpectralEstimation} shows an even stronger \input{figures/results}dependence on the SNR, with its best performance around $10$ dB and increasing errors toward both lower and higher SNRs. This may result from prominent spectral components, particularly around zero Doppler, shifting the center of mass from the velocity-related peak. 

Figure \ref{fig:reconstruction_mae} evaluates the robustness and quality of the predicted error bound $\hat{\delta}_\ell$ of our best-performing model under different SNRs. Again, for each SNR, Gaussian noise is added to the entire $\mathcal{D}^\mathrm{test}$. The primary y-axis (green numbers) shows the percentage of frames for which the absolute velocity error holds $|\hat{v}_\ell-v_\ell|\leq\hat{\delta}_\ell$, while the secondary y-axis (magenta numbers) reports the mean predicted error bound $\hat{\delta}^\mathrm{mean}$ and half of its standard deviation using our best-performing model. 
Across all SNRs, the proportion of predictions within the estimated error bound, which was set to 95\% in training, remains almost constant between $94\%$ and $95\%$, demonstrating stable error-bound estimation even under noisy conditions. The only outlier is at $-10$ dB with $91\%$, an SNR not seen in training. The mean predicted error bound is approximately $0.38$ m/s at $15$ dB to $20$ dB and increases to about $0.90$ m/s at $-10$ dB. A comparatively large standard deviation, increasing with higher noise levels, is also observed. This is desirable in this context, as it indicates that the predicted error bound adapts to individual frames rather than remaining constant. \textit{Thus, error bound $\hat{\delta}_\ell$ provides frame-dependent information about the confidence of the velocity estimate and may additionally reflect variations in signal quality.}

Figure \ref{fig:sample_error_bound} shows consecutive frames from the domain of $\mathcal{D}^\mathrm{test}$ at an SNR of 5 dB. In addition to the per-frame velocity error and predicted error bound, the velocity error is smoothed using a first-order IIR filter with coefficient $\gamma=0.9$ and an FIR filter with a non-causal rectangular impulse response of length $L'=9$. The velocity error and the predicted error bound exhibit a closely matching temporal behavior. The predicted bound is violated at only two frames ($\ell=16$ and $\ell=34$), i.e., in 5\% of the 40 displayed frames -- adhering to our chosen $\tau=0.95$. Both smoothing approaches reduce the magnitude of the velocity error, but introduce some lag.\input{figures/results_2}\input{figures/results_3}

%% file: figures/results.tex
\begin{figure}[t]
    \centering
\begin{tikzpicture}

\begin{axis}[
    name=mainaxis,
    width=0.46\textwidth,
    height=3.6cm,
    xlabel={SNR (dB)},
    ylabel={MAE (m/s)},
    ylabel style={yshift=-2mm},
    xmin=-10.2, xmax=20.2,
    ymin=0, ymax=1.5,
    grid=both,
    legend pos=north east,
    legend cell align=left,
    legend style={draw=none, fill=white,xshift=1.8mm, legend columns=2,
    yshift=0.57mm,inner sep=1pt, cells={align=left},  inner xsep=2pt,
    inner ysep=-1pt},
    tick label style={font=\small},
    label style={font=\small},
]

% CMA
\addplot[
    thick,
    draw=red,
    mark=square*,
    mark options={
        draw=red,
        fill=red
    },
    error bars/.cd,
        y dir=both,
        y explicit,
        error bar style={draw=red}
] coordinates {
    (-10,25.07)
    (-5,4.1)
    (0,0.8)  
    (5,0.49)  
    (10,0.47)  
    (15,0.53) 
    (20,0.64)
};
\addlegendentry{\texttt{CMA} \cite{Shariff2017AdaptiveSpectralEstimation}}

% Peak Picking
\addplot[
    thick,
    draw=orange,
    mark=triangle*,
    mark options={
        draw=orange,
        fill=orange
    },
    error bars/.cd,
        y dir=both,
        y explicit,
        error bar style={draw=orange}
] coordinates {
    (-10,1.17)
    (-5,0.6) 
    (0,0.42)   
    (5,0.42)  
    (10,0.42) 
    (15,0.42)  
    (20,0.42) 
};
\addlegendentry{\texttt{PP} \cite{richards2010principles}}

% Spectral Centroid
\addplot[
    thick,
    draw=blue,
    mark=square*,
    mark options={
        draw=blue,
        fill=blue
    },
    error bars/.cd,
        y dir=both,
        y explicit,
        error bar style={draw=blue}
] coordinates {
    (-10,0.37)
    (-5,0.35)
    (0,0.34)  
    (5,0.34)  
    (10,0.34)  
    (15,0.34) 
    (20,0.34)
};
\addlegendentry{\texttt{XCA} \cite{Shariff2017AdaptiveSpectralEstimation}}

% Our Best
\addplot[
    thick,
    draw=green!60!black,
    mark=*,
    mark options={
        draw=green!60!black,
        fill=green!60!black
    },
    error bars/.cd,
        y dir=both,
        y explicit,
        error bar style={draw=green!60!black}
] coordinates {
    (-10,0.36)
    (-5,0.21) 
    (0,0.20)   
    (5,0.20)   
    (10,0.17)  
    (15,0.16)  
    (20,0.16)
};
\addlegendentry{Ours w/ noise \\ $\alpha=1.0$}

\end{axis}
\end{tikzpicture}\vspace*{-2mm}
    \caption{\textbf{Velocity MAE results on noisy} $\mathcal{D}^\mathrm{test}$.}
    \label{fig:velocity}
    \vspace*{-2mm}
\end{figure}
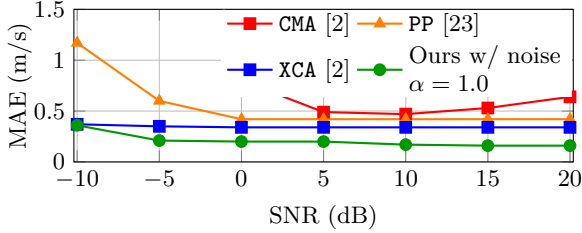

%% file: figures/results_2.tex
\begin{figure}[t]
    \centering
    \begin{tikzpicture}
        \begin{axis}[
            name=mainaxis,
            width=0.45\textwidth,
            height=3.6cm,
            xlabel={SNR (dB)},
            ylabel={},
            ylabel style={yshift=-2mm},
            yticklabel={
        \pgfmathparse{100*\tick}
        \pgfmathprintnumber{\pgfmathresult}\%
    },yticklabel style={text=green!60!black},
ylabel style={text=green!60!black},
            xmin=-10.2, xmax=20.2,
            ymin=0.8, ymax=1,
            grid=both,
            legend pos=north east,
            legend cell align=left,
            legend style={draw=none, fill=white,xshift=-2mm,
            yshift=-1mm,inner sep=1pt, cells={align=left}},
            tick label style={font=\small},
            label style={font=\small},
        ]

        % Reference line y = x
        \addplot[
            black,
            dashed,
            domain=-15.2:20.2,
            samples=2
        ] {0.95};

        % Our Best
        \addplot[
            thick,
            draw=green!60!black,
            mark=*,
            mark options={
                draw=green!60!black,
                fill=green!60!black
            },
            error bars/.cd,
                y dir=both,
                y explicit,
                error bar style={draw=green!60!black}
        ] coordinates {
            (-10, 0.906)
            (-5,0.956) 
            (0,0.953)   
            (5,0.941)   
            (10,0.943)  
            (15,0.9399)  
            (20,0.946)
        };
        
        \node at (axis cs:-3.7,0.974) {$\tau=0.95$ (\ref{eq:pinball_loss})};
        \node at (axis cs:12,0.913) {\textcolor{green!60!black}{\% of $|\hat{v}_\ell-v_\ell|\leq\hat{\delta}_\ell$}};
        \addplot[
    domain=180:90,
    samples=100,
    thick
]
({-8 + 1.0*cos(x)},
 {0.953 + 0.018*sin(x)});

        \node[
    draw=black,
    fill=white,
    inner sep=1.3pt,
    align=left,
    anchor=north east
    ] at (axis cs:19.9,0.996) {Ours w/ noise $\alpha=1.0$};
        \end{axis}

\begin{axis}[
    at={(mainaxis.south west)},
    anchor=south west,
    width=0.45\textwidth,
    height=3.6cm,
    xmin=-10.2, xmax=20.2,
    ymin=0.15, ymax=0.92,
    axis x line=none,
    axis y line*=right,
    ylabel={\textcolor{magenta}{$\hat{\delta}^\mathrm{mean}$ (m/s)}},
    ylabel style={yshift=2mm},
    yticklabel style={text=magenta},
    ylabel style={text=magenta},
    tick label style={font=\small},
    label style={font=\small},
    xtick={-10,-5,0,5,10,15,20},
    legend pos=north east,
    legend cell align=left,
    legend style={draw=none, fill=white},
]

\addplot[
    only marks,
    mark=*,
    magenta,
    mark options={draw=magenta, fill=magenta},
    error bars/.cd,
    y dir=both,
    y explicit,
]
coordinates {
    (-10, 0.90) +- (0, 1)
    (-5, 0.54) +- (0, 0.29)
    ( 0, 0.49) +- (0, 0.175)
    ( 5, 0.46) +- (0, 0.135)
    (10, 0.39) +- (0, 0.095)
    (15, 0.38) +- (0, 0.08)
    (20, 0.38) +- (0, 0.08)
};
\node at (axis cs:7.4,0.255) {\textcolor{magenta}{$\hat{\delta}^\mathrm{mean} \pm \frac{1}{2}\sigma$ }};

\end{axis}

    \end{tikzpicture}
    %\captionsetup[figure]{skip=2pt}
    \vspace*{-2mm}
    \caption{\textbf{Our method: Percentage of velocity errors below error bound} $\hat{\delta}_\ell$ \textbf{and mean error bound }$\hat{\delta}^\mathrm{mean}$ on noisy $\mathcal{D}^\mathrm{test}$; $\hat{\delta}^\mathrm{mean} \pm \frac{1}{2}\sigma$ are shown for visualization purposes.}
    \label{fig:reconstruction_mae}
\end{figure}
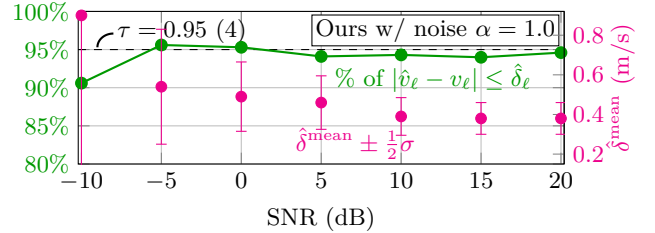

%% file: figures/results_3.tex
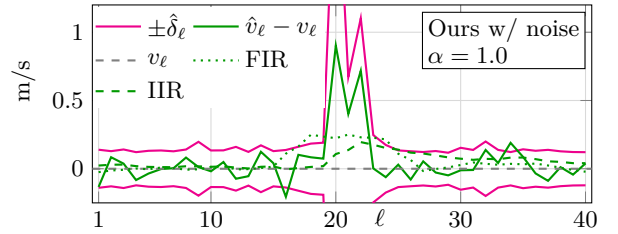
\begin{figure}[t]
    \centering
    \begin{tikzpicture}
        \begin{axis}[
            width=0.46\textwidth,
            height=4.2cm,
            xmin=0.5, xmax=40.5,
            ymin=-0.25, ymax=1.2,
            xlabel={$\ell$},
            ylabel={m/s},
            xtick={1,10,20,30,40},
            title style={
                font=\small,
                align=center
            },
            xlabel style={xshift=5mm, yshift=4.5mm},
ylabel style={xshift=3mm},
            grid=both,
            major grid style={draw=gray!30},
            minor grid style={draw=gray!15},
        legend pos=north west,
            legend cell align=left,
            legend style={draw=none, fill=white,xshift=-1.9mm, legend columns=2,
            yshift=0.6mm,inner sep=0.4pt, cells={align=left}},
            tick label style={font=\small},
            label style={font=\small},
        ]

        % Reference line y = x

% bound neg
\addplot[
    magenta,
    thick,
] 
coordinates {
(1,-0.139450)
(2,-0.128800)
(3,-0.141208)
(4,-0.121548)
(5,-0.131496)
(6,-0.133871)
(7,-0.127761)
(8,-0.140896)
(9,-0.197267)
(10,-0.133941)
(11,-0.134616)
(12,-0.160191)
(13,-0.123349)
(14,-0.171556)
(15,-0.129062)
(16,-0.137506)
(17,-0.161019)
(18,-0.184141)
(19,-0.190650)
(20,-2.442235)
(21,-0.672073)
(22,-1.102456)
(23,-0.246708)
(24,-0.174383)
(25,-0.134457)
(26,-0.128161)
(27,-0.122566)
(28,-0.129809)
(29,-0.117276)
(30,-0.153590)
(31,-0.137778)
(32,-0.199543)
(33,-0.138076)
(34,-0.122164)
(35,-0.142485)
(36,-0.131647)
(37,-0.138485)
(38,-0.127712)
(39,-0.122452)
(40,-0.121377)
};
\addlegendentry{$\pm\hat{\delta}_\ell$}

% bound pos
\addplot[
    magenta,
    thick,
    forget plot
] coordinates {
(1,0.139450)
(2,0.128800)
(3,0.141208)
(4,0.121548)
(5,0.131496)
(6,0.133871)
(7,0.127761)
(8,0.140896)
(9,0.197267)
(10,0.133941)
(11,0.134616)
(12,0.160191)
(13,0.123349)
(14,0.171556)
(15,0.129062)
(16,0.137506)
(17,0.161019)
(18,0.184141)
(19,0.190650)
(20,2.442235)
(21,0.672073)
(22,1.102456)
(23,0.246708)
(24,0.174383)
(25,0.134457)
(26,0.128161)
(27,0.122566)
(28,0.129809)
(29,0.117276)
(30,0.153590)
(31,0.137778)
(32,0.199543)
(33,0.138076)
(34,0.122164)
(35,0.142485)
(36,0.131647)
(37,0.138485)
(38,0.127712)
(39,0.122452)
(40,0.121377)
};

                % True absolute velocity error
        \addplot[
            green!60!black,
            thick,
]
 coordinates {
(1,-0.133188)
(2,0.084380)
(3,0.037493)
(4,-0.082868)
(5,-0.035010)
(6,-0.002188)
(7,0.003886)
(8,0.081885)
(9,0.012185)
(10,-0.018595)
(11,0.049295)
(12,-0.073215)
(13,-0.054501)
(14,0.124302)
(15,0.040707)
(16,-0.205550)
(17,0.101913)
(18,0.084009)
(19,0.074113)
(20,0.902392)
(21,0.401023)
(22,0.717893)
(23,0.002462)
(24,-0.061616)
(25,0.029932)
(26,-0.079070)
(27,0.053891)
(28,-0.027722)
(29,-0.048377)
(30,-0.011289)
(31,-0.009112)
(32,0.137735)
(33,0.042236)
(34,0.189771)
(35,0.086547)
(36,-0.034255)
(37,-0.088580)
(38,0.011501)
(39,-0.043002)
(40,0.038892)
};
        \addlegendentry{$\hat{v}_\ell-v_\ell$}

        \addplot[
            gray,
            dashed,
            thick,
            domain=0.5:40.5,
            samples=2
        ] {0};
        \addlegendentry{$v_\ell$}

% rect9
\addplot[green!60!black,
            thick, dotted] coordinates {
(1,-0.024690)
(2,-0.013138)
(3,-0.007408)
(4,0.001544)
(5,-0.003714)
(6,0.009019)
(7,0.005120)
(8,-0.007180)
(9,-0.004029)
(10,0.013673)
(11,0.018439)
(12,-0.004832)
(13,-0.002606)
(14,0.005374)
(15,0.015675)
(16,0.110463)
(17,0.163157)
(18,0.248978)
(19,0.235440)
(20,0.224071)
(21,0.250236)
(22,0.230126)
(23,0.226780)
(24,0.215465)
(25,0.109824)
(26,0.064012)
(27,-0.016767)
(28,-0.001737)
(29,0.009803)
(30,0.027562)
(31,0.045964)
(32,0.036170)
(33,0.029408)
(34,0.036062)
(35,0.032538)
(36,0.037872)
(37,0.023826)
(38,0.007997)
(39,-0.009441)
(40,-0.021716)
};
\addlegendentry{FIR}

% iir
\addplot[green!60!black,
            thick, dashed] 
coordinates {
(1,0.023538)
(2,0.029622)
(3,0.030410)
(4,0.019082)
(5,0.013673)
(6,0.012087)
(7,0.011267)
(8,0.018328)
(9,0.017714)
(10,0.014083)
(11,0.017604)
(12,0.008522)
(13,0.002220)
(14,0.014428)
(15,0.017056)
(16,-0.005204)
(17,0.005507)
(18,0.013357)
(19,0.019433)
(20,0.107729)
(21,0.137058)
(22,0.195142)
(23,0.175874)
(24,0.152125)
(25,0.139906)
(26,0.118008)
(27,0.111596)
(28,0.097664)
(29,0.083060)
(30,0.073625)
(31,0.065352)
(32,0.072590)
(33,0.069555)
(34,0.081576)
(35,0.082073)
(36,0.070440)
(37,0.054538)
(38,0.050235)
(39,0.040911)
(40,0.040709)
};
\addlegendentry{IIR}

        \node[
    draw=black,
    fill=white,
    inner sep=2pt,
    align=left,
    anchor=north east
    ] at (axis cs:40,1.15) {Ours w/ noise\\$\alpha=1.0$};
        \end{axis}

    \end{tikzpicture}\vspace*{-2mm}
    \caption{\textbf{Per-frame velocity error and predicted error bound}: Sequence from $\mathcal{D}^\mathrm{test}$ at an SNR of 5 dB, including the velocity error smoothed using an IIR and FIR filter.}
    \label{fig:sample_error_bound}
\end{figure}

%% file: sections/6_Conclucions.tex
\section{Conclusions}
\label{sec:conclusions}
We presented a confidence-aware speed-over-ground estimation method for Doppler RaDAR. The proposed model consistently outperforms classical baselines across all investigated SNR levels, while the predicted frame-wise error bounds reliably reflect the actual estimation error. The results confirm robustness of our approach under adverse signal conditions.
\newpage
\clearpage
\section{Acknowledgement}
The research leading to these results is funded by the German Federal Ministry for Economic Affairs and Energy within the project “CONTROL – Controlling Risk for Highly Automated Transportation Systems Operating in Complex Open Environments". The authors would like to thank the consortium for the successful cooperation.
\section{Compliance with Ethical Standards}
This study uses real RaDAR measurement data acquired by a sensor mounted underneath a train and directed toward the track bed. No human or animal subjects were involved in the study, and therefore no ethical approval was required.